%% file: main.tex
\documentclass[conference]{IEEEtran}
\IEEEoverridecommandlockouts
\usepackage{amsmath,amssymb,amsfonts}
\usepackage{amsthm} 
\usepackage{algorithmic}
\usepackage{graphicx}
\usepackage{textcomp}
\usepackage{xcolor}
\usepackage{dsfont}

\usepackage[numbers,sort&compress]{natbib}

\DeclareMathOperator*{\argmin}{\arg\!\min}

\newif\ifarxiv
\arxivtrue

\def\BibTeX{{\rm B\kern-.05em{\sc i\kern-.025em b}\kern-.08em
    T\kern-.1667em\lower.7ex\hbox{E}\kern-.125emX}}
\begin{document}

\include{defs}

\title{
Sketching and Sequence Alignment: \\
A Rate-Distortion Perspective
}

\author{\IEEEauthorblockN{Ilan Shomorony}
\IEEEauthorblockA{
University of Illinois at Urbana-Champaign\\
Urbana, IL, USA \\
ilans@illinois.edu}
\and
\IEEEauthorblockN{Govinda M. Kamath}
\IEEEauthorblockA{
Microsoft Research New England\\
Cambridge, MA, USA \\
gokamath@microsoft.com}
}


\ifarxiv
\onecolumn
\fi

\maketitle

\begin{abstract}
Pairwise alignment of DNA sequencing data is a ubiquitous task in bioinformatics and typically represents a heavy computational burden. 
A standard approach to speed up this task is to compute ``sketches'' of the DNA reads (typically via hashing-based techniques) that allow the efficient computation of pairwise alignment scores.
We propose a 
rate-distortion framework to study the problem 
of computing sketches that achieve the optimal tradeoff between sketch size and alignment estimation distortion.
We consider the simple setting of i.i.d.~error-free sources of length $n$ and introduce a new sketching algorithm called ``locational hashing.'' 
While standard approaches in the literature based on min-hashes require $B = (1/D) \cdot O\left( \log n \right)$ bits to achieve a distortion $D$,
our proposed approach only requires $B = \log^2(1/D) \cdot O(1)$ bits.
This can lead to significant computational savings in pairwise alignment estimation.
%
%
%
\end{abstract}




\section{Introduction}
\input{intro}

\section{Problem Setting} \label{sec:problem}
\input{problem}

\section{Min-Hash-based Sketching} \label{sec:minhash}
\input{minhash}

\section{Main Results} \label{sec:result}
\input{result}

\ifarxiv

\section{Extensions to the Noisy Setting} \label{sec:noisy}
\input{noisy}

\fi

\section{Proof of Main Result} 
\label{sec:proof}
\input{proof}

\section{Discussion} \label{sec:discussion}
\input{discussion}

\vspace{5mm}

\ifdraft


\fi



\newpage

{
\bibliographystyle{ieeetr}
\bibliography{refs.bib}
}

\newpage

\onecolumn
\appendix

\input{appendix}

\end{document}

%% file: defs.tex
\newtheorem{claim}{Claim}
\newtheorem{corollary}{Corollary}
\newtheorem{theorem}{Theorem}
\newtheorem{defn}{Definition}
\newtheorem{example}{Example}
\newtheorem{lemma}{Lemma}
\newtheorem{proposition}{Proposition}
\newtheorem{remark}{Remark}


\renewcommand{\L}{{\mathcal L}}

\newcommand{\1}{{\bf 1}}
\newcommand{\Z}{{\mathds Z}}
\newcommand{\dis}{{\mathsf{dis}} \,}
\newcommand{\ep}{\epsilon}
\newcommand{\vep}{\varepsilon}

\newcommand{\bA}{\mathsf{A}}
\newcommand{\bC}{\mathsf{C}}
\newcommand{\bG}{\mathsf{G}}
\newcommand{\bT}{\mathsf{T}}

\newcommand{\A}{{\mathcal A}}
\newcommand{\B}{{\mathcal B}}
\newcommand{\C}{{\mathcal C}}

\newcommand{\G}{{\mathcal G}}
\renewcommand{\H}{{\mathcal H}}
\newcommand{\D}{{\mathcal D}}

\newcommand{\Dr}{\mathcal{DR}}
\newcommand{\dof}{\mathbf{D}}
\newcommand{\Dreg}{\dof}
\newcommand{\Enc}{{\rm Enc}}
\newcommand{\Dec}{{\rm Dec}}
\newcommand{\Sk}{{\rm Sk}}
\newcommand{\minhash}{h^{\min }}

\newcommand{\E}{{\mathcal E}}
\newcommand{\Fs}{{\mathcal F}}
\newcommand{\V}{{\mathcal V}}
\renewcommand{\S}{{\mathcal S}}
\newcommand{\I}{{\mathcal I}}
\newcommand{\M}{{\mathcal M}}
\newcommand{\N}{{\mathcal N}}
\newcommand{\U}{{\mathcal U}}
\newcommand{\T}{{\mathcal T}}
\newcommand{\IN}{{\mathbb N}}
\newcommand{\R}{{\mathbb R}}
\newcommand{\Rs}{{\mathcal R}}
\newcommand{\Os}{{\mathcal O}}
\newcommand{\Ps}{{\mathcal P}}
\newcommand{\K}{{\mathcal K}}
\newcommand{\W}{{\mathcal W}}
\newcommand{\X}{{\mathcal X}}
\newcommand{\vX}{{\vec{X}}}
\newcommand{\Y}{{\mathcal Y}}
\newcommand{\vY}{{\vec{Y}}}
\newcommand{\F}{{\mathbb F}}
\newcommand{\dE}{D_\Sigma}
\newcommand{\q}[2]{Q_{s_{#1},d_{#2}}}
\newcommand{\p}[2]{P_{s_{#1},d_{#2}}}
\newcommand{\m}[2]{M_{s_{#1},d_{#2}}}
\newcommand{\ttt}{3 \times 3 \times 3}
\newcommand{\kkk}{K \times K \times K}
\newcommand{\kk}[1]{#1 \times #1 \times #1}
\newcommand{\cT}{{\cal T}}
\newcommand{\cR}{{\cal R}}
\newcommand{\cN}{{\cal N}}
\newcommand{\cC}{{\cal C}}

\newcommand{\s}{{\bf s}}
\newcommand{\bs}{{\bf s}}
\newcommand{\bc}{{\bf c}}

\newcommand{\EP}{{\rm EP}}

\newcommand{\setx}{\{ x_{(i)}^{K} \}_M }
\newcommand{\setxM}[1]{\{ x_{(i)}^{K} \}_{#1} }

\newcommand{\setX}{\{ X_{(i)}^{K} \}_M }
\newcommand{\setXM}[1]{\{ X_{(i)}^{K} \}_{#1} }

\newcommand{\sety}{\{ y_{(i)}^{K} \}_N }
\newcommand{\setyN}[1]{\{ y_{(i)}^{K} \}_{#1} }

\newcommand{\setY}{\{ Y_{(i)}^{K} \}_N }
\newcommand{\setYN}[1]{\{ Y_{(i)}^{K} \}_{#1} }

\newcommand{\bp}{{\bf p}}
\renewcommand{\r}{{\bf r}}
\newcommand{\x}{{\bf x}}
\newcommand{\y}{{\bf y}}
\newcommand{\z}{{\bf z}}

\newcommand{\Cunc}{C_\text{unc}}

\newcommand{\aln}[1]{\begin{align*}#1\end{align*}}

\newcommand{\al}[1]{\begin{align}#1\end{align}}

\newcommand{\js}{{\rm JS}}
\newcommand{\sjs}{{\rm SJS}}

\newcommand{\one}{\mathds{1}}

\renewcommand{\1}{\mathbf{1}}

\newcommand\Ber{{\rm Ber}}
\newcommand\plog{\text{polylog}}

\renewcommand{\top}{T}
\newcommand{\xor}{\oplus}

\newcommand{\uvec}{{\mathbf u}}
\newcommand{\vvec}{{\mathbf v}}

\renewcommand{\u}{\uvec}
\renewcommand{\v}{\vvec}

\renewcommand{\p}{\mathbf{p}}
\renewcommand{\q}{\mathbf{q}}

\newcounter{numcount}
\setcounter{numcount}{1}

\newcommand{\eqnum}{\stackrel{(\roman{numcount})}{=}\stepcounter{numcount}}
\newcommand{\leqnum}{\stackrel{(\roman{numcount})}{\leq\;}\stepcounter{numcount}}
\newcommand{\geqnum}{\stackrel{(\roman{numcount})}{\geq\;}\stepcounter{numcount}}
\newcommand{\cnt}{$(\roman{numcount})$ \stepcounter{numcount}}
\newcommand{\rescnt}{\setcounter{numcount}{1}}

\newif\iflong
\longtrue

\newif\ifdraft
\drafttrue

\newcommand{\iscomment}[1]{
\ifdraft
{\color{blue} \bf{{{{IS --- #1}}}}}
\else
\fi
}    

\newcommand{\gkcomment}[1]{
\ifdraft
{\color{red} \bf{{{{gmk --- #1}}}}}
\else
\fi
}    

%% file: intro.tex

A key step in many bioinformatics analysis pipelines is the identification of regions of
similarity between pairs of sequences (typically DNA sequencing reads).
This task, known as (pairwise) sequence
alignment, is a heavy computational burden,
particularly in the context of sequencing data from long-read sequencing technologies such as PacBio and Oxford Nanopore.
These technologies typically produce DNA reads of lengths in the tens of thousands,
but at the expense of significantly higher error rates ($\sim\hspace{-1mm} 10\%$)
\cite{pacbio_errorRate}.

\begin{figure}[b] 
\vspace{-4mm}
\centering
\ifarxiv
\includegraphics[width = .4\columnwidth]{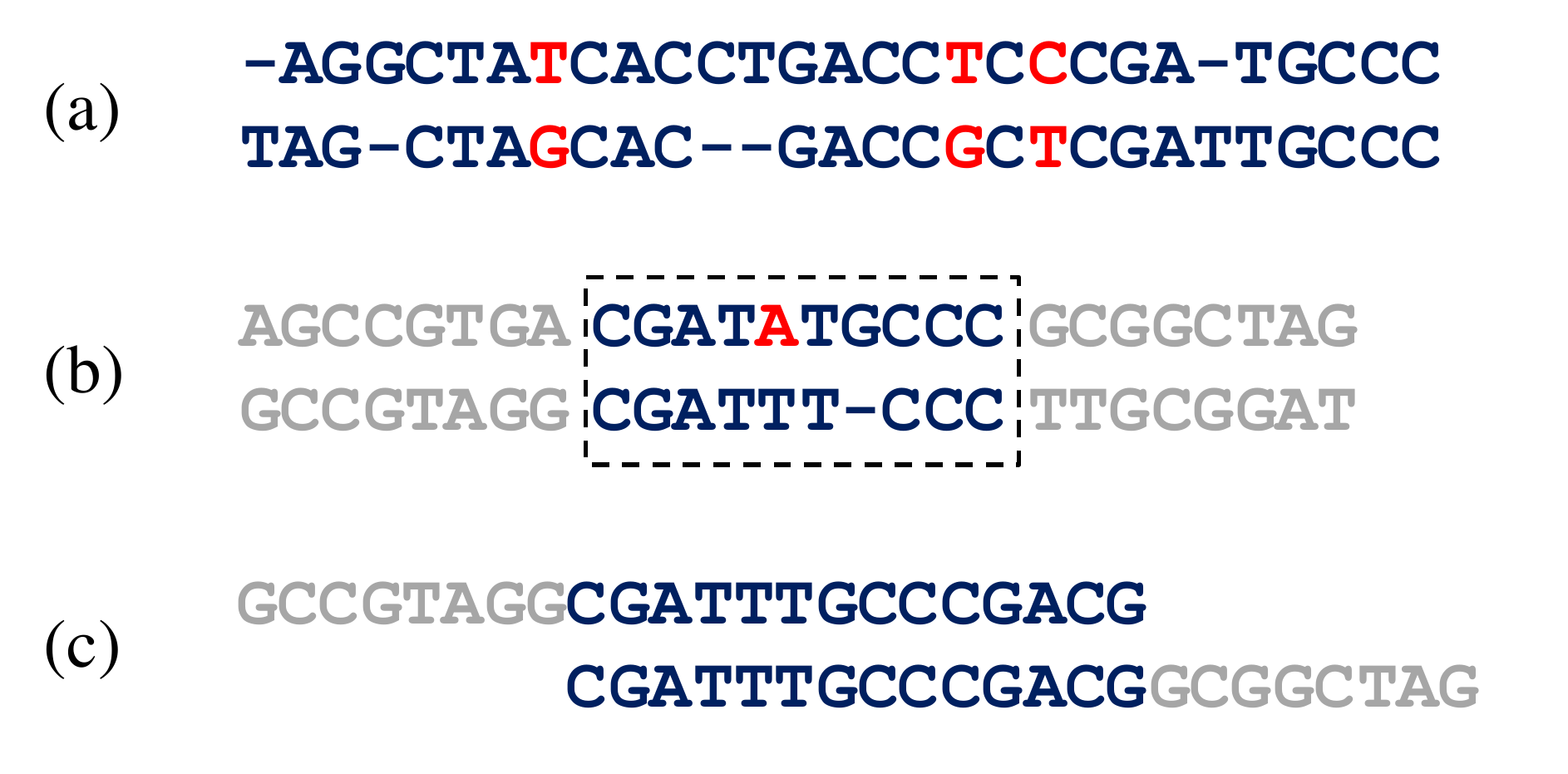}
\fi
\vspace{-2mm}
    \caption{(a) Optimal \emph{global} alignment between two sequences. Dashes represent indels while red symbols represent mismatches. 
    The optimal alignment maximizes a score of the form $M \times \text{matches} - D \times \text{dashes} - S \times \text{mismatches}$ \cite{SmithWaterman}.
    (b) In a \emph{local} alignment, only an interior segment of the sequences is required to be aligned. (c) An (error-free) overlap is a special kind of alignment where a suffix of one sequence should be aligned to a prefix of the other.
    \label{fig:alignment}}
\end{figure}

The optimal alignment between two sequences of length $n$, illustrated in Figure~\ref{fig:alignment}, can be found via dynamic programming in $O(n^2)$ time \cite{SmithWaterman}.
Hence, the time complexity for the pairwise alignment of $m$ reads of length $n$ is $O(m^2 n^2)$, which is impractical even for bacterial genome datasets (for which we may have $m \sim 10^5$ long reads of length $n \sim 10^4$).
This challenge is commonly addressed 
by first converting each long DNA sequence into a short \emph{sketch} of length $s$. The sketch can be thought of as low-dimensional representation of the original sequence, and is typically constructed using hashing techniques.
The sketching procedure is designed in a way that estimating the alignment score between two sequences from their sketches is straightforward and can be performed in $O(s)$ time.
Based on this idea, most practical sequence alignment pipelines are based on a two-step approach:
they first use pairwise sketch comparisons (which take $O(m^2s)$ time), and then perform more careful alignment only on a small set of selected pairs that are ``likely'' to have a significant alignment
\cite{Myers2014,Berlin2015,li2016minimap,li2018minimap2, ondov2016mash, jain2017fast}.
This approach can be cast under the more general framework of sketching algorithms \cite{indyk2007sketching}, which are becoming increasingly popular in genomic data science \cite{sketching_review,marcais_sketching,sjs_cell, kamath2020adaptive}.

At a high level, the goal of sketching-based sequence alignment is to individually \emph{compress} sequences
down to (typically sublinear-size) sketches, from which pairwise alignments can be accurately estimated.
This suggests a natural tradeoff between compression level and alignment accuracy.
Motivated by this, we propose a rate-distortion framework to study this fundamental tradeoff and characterize optimal sketching.

\begin{figure}[t] 
\centering
\ifarxiv
\includegraphics[width = .4\columnwidth]{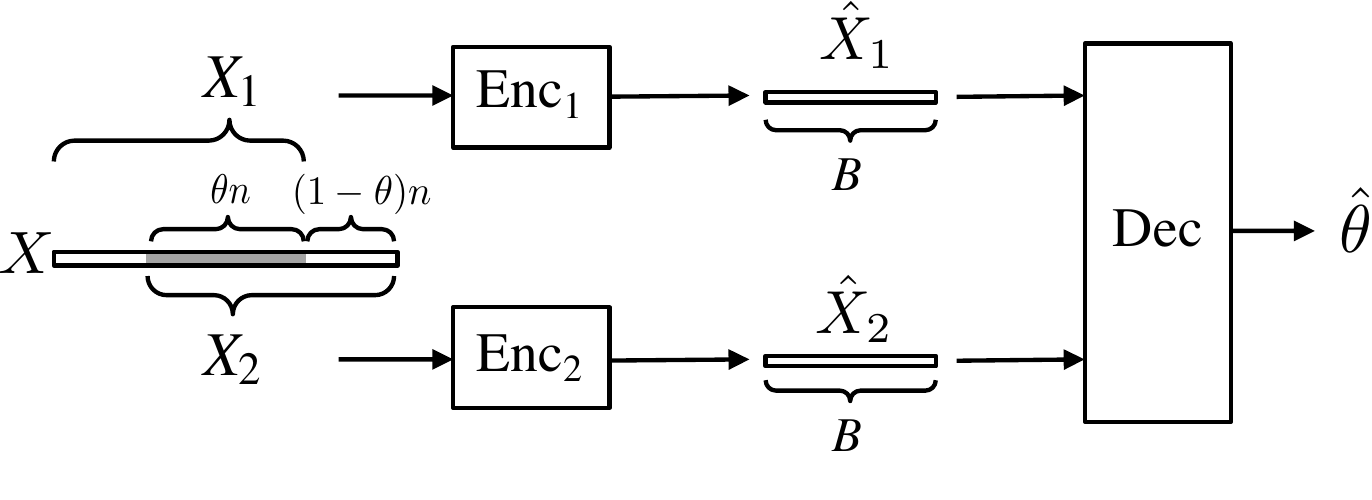}
\fi
    \caption{The source sequences $X_1$ and $X_2$ are individually encoded into $B$-bit sketches, from which the overlap $\theta$ must be estimated.
    \label{fig:sourcecoding}}
\vspace{-3mm}
\end{figure}

We study the problem under a simplified setting where we have two binary length-$n$ sequences $X_1$ and $X_2$.
Moreover,
we focus on a specific type of alignment -- an error-free \emph{overlap} (see Figure~\ref{fig:alignment}(c)) -- since overlaps are the main type of alignment one is interested in in the context of genome assembly \cite{HINGE}.
We assume that $X_1$ and $X_2$ have an overlap of size $\theta n$ and consider a distributed source coding formulation, where $X_1$ and $X_2$ are compressed down to $B$ bits each,
as illustrated in Figure~\ref{fig:sourcecoding}.
From the $B$-bit sketches, we would like to estimate $\theta$ under a quadratic distortion constraint
\al{
E[(\theta - \hat \theta)^2] \leq D.
\label{eq:dist}
}
We are interested in characterizing the fundamental tradeoff between the pairs $(B,D) \in \R_+^2$ that can be achieved in the asymptotic regime $n \to \infty$.


As we discuss in Section~\ref{sec:minhash}, existing sketching algorithms based on min-hashes \cite{broder1997syntactic_minHash} can achieve a bounded distortion $D$ with sketches of size $B =  (1/D) \cdot O(\log n)$ bits.
Our main result surprisingly shows that it is possible to improve over that and achieve a bounded distortion $D$ with sketches of a constant (in $n$) length $B = \log^2 \left(1/D \right) \cdot \Theta(1)$.
To this end, we introduce a new sketching algorithm, which we term \emph{locational hashing}.
In addition, we show that a lower bound on the sketch size needed is $B = \Omega(\log(1/D))$, implying that the locational hashing scheme is not far from the optimum.

Finally, we notice that, from a practical standpoint, sketching procedures with linear-time encoding and decoding  are desirable.
Otherwise, the two-step approach to alignment outlined above is ineffective.
While all schemes discussed in this paper do have the property of linear-time encoding and decoding, we do not formally place this as a constraint for the source coding problem shown in Figure~\ref{fig:sourcecoding}.

%% file: problem.tex

The distributed source coding problem \cite{slepianwolf,AhlswedeKorner,WynerSide,WynerZiv} 
is concerned with the compression of correlated sources by encoders that do not communicate with each other.
Our problem formulation is related to 
the literature on distributed hypothesis testing and parameter estimation \cite{ahlswede1986hypothesis,amari1998statistical}.
We assume that 
two length-$n$ source sequences $X_1$ and $X_2$ have an overlap of size $\theta n$. 
We assume that the unknown overlap parameter satisfies $\theta \in \{0\} \cup [\theta_0,1]$ for some fixed  $\theta_0 > 0$.
This is because typically in practice one is not interested in accurately estimating small overlaps, only no-overlaps or moderately sized overlaps \cite{kamath2020adaptive}.

Our goal is to design an encoder (i.e., a sketching procedure) with small encoding rates and a decoder that takes the sketches and attempts to estimate the (normalized) overlap size $\theta$.
We focus on a quadratic distortion metric, thus 
seeking a decoder
that produces an estimator $\hat \theta$ 
satisfying (\ref{eq:dist}) for some $D$.

For a sequence length $n$, the joint distribution $(X_1,X_2)$ is defined as follows.
First, a single i.i.d.~$\Ber(1/2)$ sequence $X$ of length $(2-\theta)n$ is generated.
In the context of DNA sequencing, $X$ can be thought of as the underlying genome sequence.
We then define $X_1 = X[1:n]$ (the first $n$ bits of $X$) and $X_2 = X[(1-\theta)n + 1: (2-\theta)n]$ (the last $n$ bits of $X$).
Notice that the length-$\theta n$ suffix of $X_1$ matches the length-$\theta n$ prefix of $X_2$.
In this paper, we assume that $X_1$ and $X_2$ are not corrupted by noise.
Extensions to the more realistic noisy setting are briefly discussed in Section~\ref{sec:discussion}.

We will also define the \emph{overlap fraction} $\alpha$ to be the ratio between the overlap size $\theta n$ and the length of $X$; i.e.,
\al{
\alpha \triangleq \frac{\theta n}{(2-\theta)n} = \frac{\theta}{2-\theta}. 
\label{eq:alpha}
}
Notice that $\alpha \in (0,1)$ and $\theta$ are in one-to-one correspondence.

Unlike in standard source coding, where a length-$n$ source $X_i$ is compressed into $n R$ bits, for a positive rate $R$, the regime of interest for our problem requires sketches of constant size in $n$.
Our main results show that this is in fact feasible.



\begin{defn}
A sketching procedure with sketch length $B$ is a mapping 
$\Sk : \{0,1\}^n \to \{0,1\}^B$.
\end{defn}

The sketching procedure is independently applied to each source, producing $\hat X_i = \Sk(X_i)$.
The decoder's task is then to estimate $\theta$ from 
$\hat X_1$ and $\hat X_2$.

\begin{defn}
A decoder is any function $\Dec : \{0,1\}^B \times \{0,1\}^B \to (0,1)$.
\end{defn}

The decoder is used to estimate $\hat \theta = \Dec(\hat X_1,\hat X_2)$.
We say that a rate-distortion pair $(B,D)$ is achievable if, for sufficiently large $n_0$, there exists a sketching procedure for any $n \geq n_0$ with sketch length $B$ satisfying 
\al{
\max_{\theta} E[(\theta - \hat \theta)^2] \leq D.
\label{eq:maxdist}
}
The rate-distortion function $B(D)$ (i.e., the minimum number of sketching bits needed to guarantee distortion $D$) is defined as $B(D) = \inf \{B : (B,D) \text{ is achievable}\}$.

\vspace{2mm}
\noindent {\bf Notation: }
Throughout we assume that $\log(\cdot)$ refers to the base-$2$ logarithm.
For $x \in \R$, we let $(x)^+ = \max(x,0)$.
For a positive integer $m$ we let $[1:m] = \{1,\dots,m\}$.
For a real number $x \in (0,1)$, we let  
$x = \sum_{i=1}^\infty x_i 2^{-i}$, $x_i \in \{0,1\}$, be its (minimal) binary expansion and for some $b \in \Z_+$, we let 
\aln{
[x]_b \triangleq \sum_{i=1}^b x_i 2^{-i}
} 
be a $b$-bit truncation of $x$.

%% file: minhash.tex

\begin{figure}[t] 
\centering
\ifarxiv
\includegraphics[width = 0.6\columnwidth]{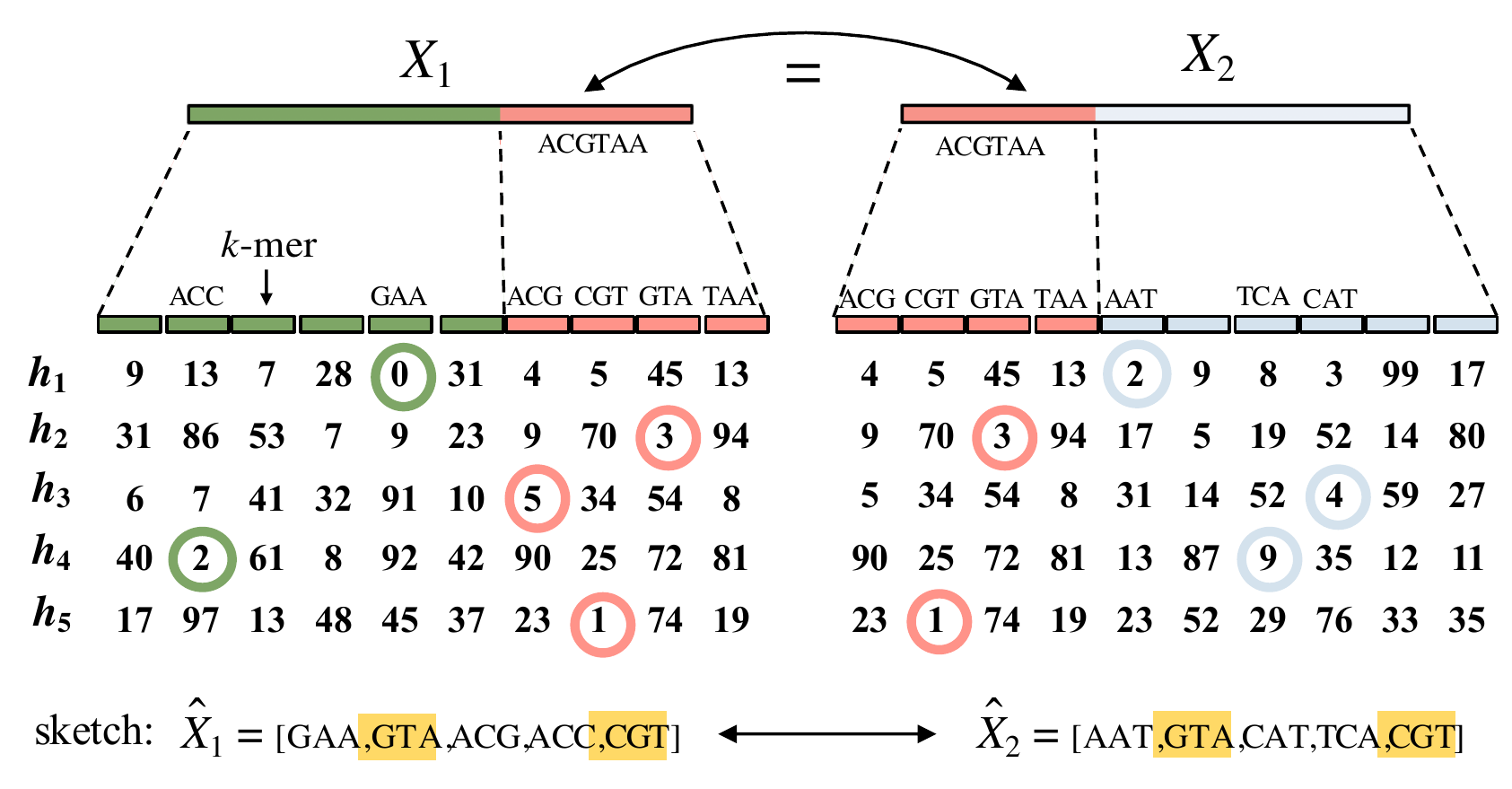}
\fi
\vspace{-3mm}
    \caption{Estimating the overlap $\theta$ between two reads using min-hash-based sketches \cite{Berlin2015} with $k=3$.
    Since there are two min-hash matches, we have estimates $\hat \alpha = 2/5$ and $\hat\theta = (4/5)/(1+2/5) = 4/7$.
    \label{fig:minhash}}
\vspace{-5mm}
\end{figure}

As a concrete example motivated by practice, we consider the min-hash-based sketching procedure from the MHAP algorithm \cite{Berlin2015}, illustrated in Figure~\ref{fig:minhash}.
For a parameter $k$, we let $\Gamma(X_i)$ be the set of all $k$-mers of $X_i$; i.e., the set of all $n-k+1$ length-$k$ substrings of $X_i$.
For a hash function $h : \{0,1\}^k \to \Z$, we define the \emph{min-hash} of $X_i$ to be
\al{
    \minhash(X_i) \triangleq
    \argmin_{x \in \Gamma(X_i)} h(x).
    \label{eq:minhash}
}
A sketch can then be defined by taking $m$ distinct hash functions $h_1,\dots,h_m$ and concatenating the min-hashes as
\al{
\Sk(X_i) = [\minhash_1(X_i),\dots,\minhash_m(X_i)].
\label{eq:mhapsketch}
}
The key motivation behind this sketching procedure
is that
for a randomly selected hash function $h$ (so that all $k$-mers are equally likely to be the minimizer of $h$), it can be shown (see, for example, \cite[Chapter 3]{leskovec2020mining}) that
\begin{align}
        \Pr\left[ \minhash(X_1) =  \minhash(X_2)\right] =
    \frac{|\Gamma(X_1) \cap \Gamma(X_2)|}{|\Gamma(X_1) \cup \Gamma(X_2)|}.
    \label{eq:prjk}
\end{align}
The right-hand side of (\ref{eq:prjk}) is known as the \emph{Jaccard similarity} between the sets $\Gamma(X_1)$ and $\Gamma(X_2)$.

Given the i.i.d.~source setting from Section~\ref{sec:problem}, we would like to choose $k$ so that $k$-mers in $X$ are unique, so that the event $\{\minhash(X_1) =  \minhash(X_2)\}$ must be caused by a $k$-mer that occurs in the overlap between $X_1$ and $X_2$.
It is straightforward to show 
that, 
if $k \geq 3 \log n$, then the probability that $X$ contains two identical $k$-mers vanishes as $n \to \infty$.
In fact, if we choose $k = 3 \log n$, with probability 
at least $1 - 1/n$,  
\al{
    \frac{|\Gamma(X_1) \cap \Gamma(X_2)|}{|\Gamma(X_1) \cup \Gamma(X_2)|} = \frac{\theta n - k + 1}{(2-\theta)n - k + 1} \overset{n \to \infty}{\longrightarrow} 
    \alpha.
}
This means that by comparing the sketches $\Sk(X_1)$ and $\Sk(X_2)$ from (\ref{eq:mhapsketch}) one entry at a time, we obtain $m$ independent Bernoulli random variables
\aln{
Z_j = \one\{ \minhash_j(X_1) =  \minhash_j(X_2) \}
}
for $j=1,\dots,m$, each having $\Pr(Z_j = 1) \to \alpha$ as $n \to \infty$.
Therefore, a natural estimator for $\alpha$ is
\al{
\hat \alpha = \frac{1}{m} \sum_{j=1}^m Z_j.
\label{eq:mhapest}
}
An estimator for $\theta$ can also be obtained by setting $\hat{\theta}(\alpha) = \frac{2\alpha}{1+\alpha}$.
From the Delta method \cite[Chapter 3]{van2000asymptotic}, we have that, as  $m \rightarrow \infty$, $\text{Var}(\hat{\theta}(\hat{\alpha})) = (\hat{\theta}'(\hat{\alpha}))^2\text{Var}(\hat{\alpha}) \le 
\text{Var}(\hat{\alpha}) [\sup_{\alpha \in [0,1]} (\hat{\theta}' 
(\alpha))^2] = 4\text{Var}(\hat{\alpha})$.
Since 
$\text{Var}(\hat{\alpha}) = \frac{\alpha(1-\alpha)}{m} \le \frac{1}{4m}$,
 we have that, for large $m$, $D \approx \text{Var}(\hat{\theta}) \le \frac{1}{m}$. 
%
%
%
We conclude that, 
to achieve a distortion $D$ 
using the 
sketching procedure in (\ref{eq:mhapsketch}),
we need
\al{
B = mk \leq m (3 \log n) \le \frac{3 \log n}{ D}
\label{eq:mhapbits}
}
bits.
Hence, while this scheme achieves a finite distortion (which can be made arbitrarily small by increasing $B$), the sketch size $B$ is logarithmic in $n$.

We conclude this section by pointing out that one could in principle modify the scheme described above to work with $k = O(1)$, but it is not straightforward  to analyze the scheme in the presence of repeated $k$-mers.
Moreover, we point out that the min-hash defined in (\ref{eq:minhash}) is more commonly defined as returning the value of the minimum hash itself instead of the minimizing $k$-mers (i.e., $h(x)$ instead of $x$).
However, in order to guarantee that all $k$-mers have a different hash value, the number of bits required per min-hash is at least $k$, so this alternative definition does not affect the analysis above.

%% file: result.tex

Our main contribution is a new sketching procedure that significantly improves over the required number of bits $B$ 
in (\ref{eq:mhapbits}).
Combined with a simple lower bound on the minimum sketch size, we have the following main result.

\begin{theorem} \label{thm:main}
The minimum sketch size $B$ required to achieve distortion $D$  satisfies
\al{
\frac18 \log \left(\frac{1-\theta_0}{D}\right) \leq B(D) 
\leq \frac{24^2}{\alpha_0^2} \ln^2 \left(\frac3D\right)
}
for any $D$ small enough.
\end{theorem}

We first introduce the new sketching technique and then analyze its performance.

\subsection{Locational Hashing} \label{sec:simpleloc}

Our sketching procedure is motivated by the min-hash approach from Section~\ref{sec:minhash}.
But instead of recording the minimizer $k$-mers of certain hash functions, we record the approximate \emph{location} of special substrings in the sequence $X_i$.

We start by introducing a simple construction that stores only one locational hash, and then improve upon it.
Suppose we have a sequence $X_1$ for which we want to compute a sketch $\hat X_1$.
We start by lexicographically sorting the set 
\aln{
\S = \{X_1 [m:n]\cdot \$ : 1 \leq m \leq n\}
}
of all suffixes of $X_1$.
To make the sorting operation well-defined, we append to each suffix $X_1 [m:n]$ a character $\$$, which is assumed to be the last character in the lexicographic order of the alphabet.
We then take the lexicographically first string from $\S$, and its corresponding location $m_1$.
Notice that, for the case of binary strings, this essentially corresponds to taking the starting point of the longest run of zeros in $X_1$
(but the lexicographic approach extends to larger alphabets).

Our sketch will be constructed by storing the first $B$ bits of the binary expansion of $m_1/n$; i.e.,
\aln{
\hat X_1
= [ m_1/n ]_B.
}
Notice that the difference between $[ m_1/n ]_B$ and $m_1/n$ is at most $2^{-B}$.
Let $m_2$ and $\hat X_2$ be the counterparts of $m_1$ and $\hat X_1$ for $X_2$.
The decoder can then estimate the overlap $\theta$ from the sketches $\hat X_1$ and $\hat X_2$ by simply setting
\al{
\hat \theta = 1- (\hat X_1 - \hat X_2)^+.
}
To analyze the performance of this estimator, consider the original length-$(2-\theta)n$ sequence $X$, from which $X_1$ and $X_2$ were generated.
Notice that, for large $n$, the location of the lexicographically first suffix of $X$ is essentially uniformly distributed over all possible starting locations.
Hence the probability that the lexicographically first suffix of $X$ falls in the overlap between $X_1$ and $X_2$ approaches
\aln{
\frac{\theta n}{(2-\theta)n} = \alpha
}
as $n \to \infty$.
This implies that, with probability approaching $\alpha$, the lexicographically first suffix of $X_1$ and $X_2$ corresponds to the same location in $X$, and we have
\al{
m_1 = m_2 + (1-\theta) n.
\label{eq:shift}
}
Thus, with a probability approaching $\alpha$, the estimator satisfies
\aln{
|\theta - \hat \theta| \leq  2\cdot 2^{-B} = 2^{-(B-1)}.
}
Whenever (\ref{eq:shift}) is not satisfied (which occurs with probability approaching $1-\alpha$ as $n \to \infty$), the absolute error of the estimator is at most $1$, which implies that
\al{
E[(\theta-\hat \theta)^2 ] \leq \alpha 2^{-2(B-1)} + (1-\alpha). \label{eq:simpleloc}
}
We see that the locational hashing approach to sketching allows us to obtain a term in the distortion bound (\ref{eq:simpleloc}) of the form $2^{-B}$.
However, because the estimator only ``works'' with probability $\alpha$, our achieved distortion cannot go below $1-\alpha$.
In the next subsection, we fix this issue by repeating the locational hashing procedure for distinct lexicographic orders.

\subsection{Improved Locational Hashing} \label{sec:multipleloc}

\begin{figure}[b] 
\vspace{-3mm}
\centering
\ifarxiv
\includegraphics[width = 0.4\columnwidth]{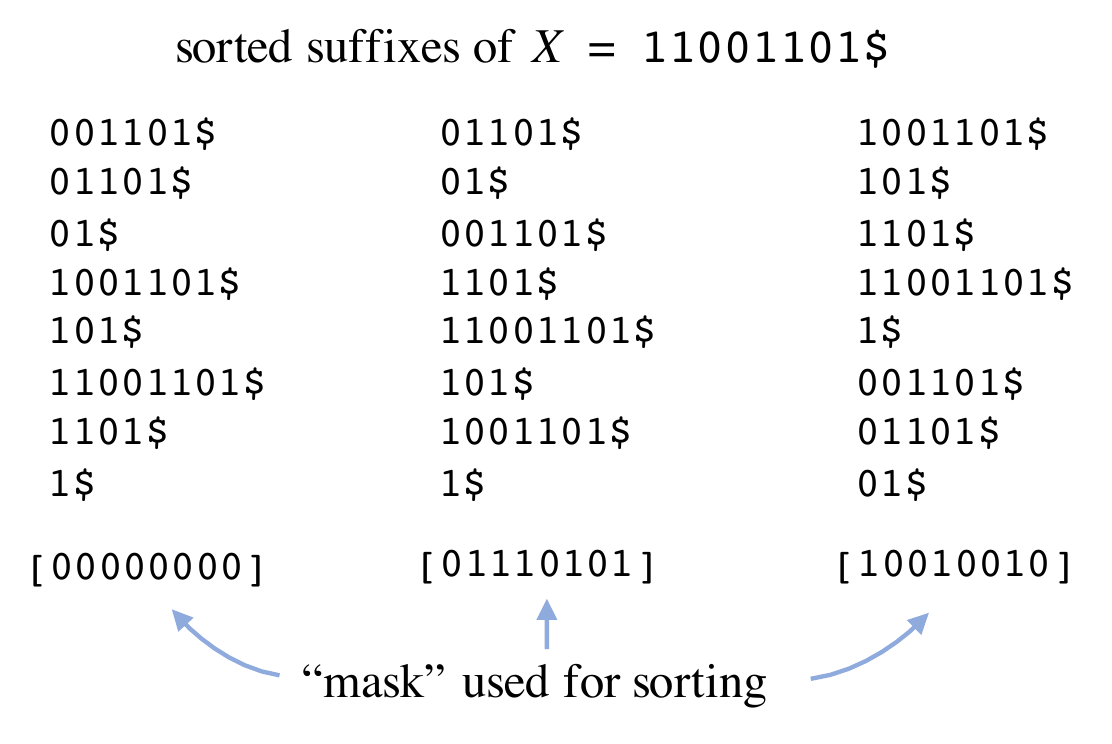}
\fi
\vspace{-1mm}
    \caption{Lexicographic sorting of the suffixes of a length-8 sequence $X$ for three different lexicographic orderings.
    The mask is a binary sequence whose $i$th entry indicates whether, at the $i$th entry, 0 or 1 should be considered to be lexicographically first.
    \$ is always the lexicographically last character.
    \label{fig:sorting}}
\end{figure}

In order to improve on the performance of the sketching procedure in Section~\ref{sec:simpleloc}, we consider using different lexicographic orders to compare strings (while maintaining \$ as the last 
character in the lexicographic order).
We do this by picking an ordering of $0$ and $1$ at each 
position uniformly at random, independent across positions. 
In other words, for position $i$ \underline{of the suffix}, $ 1 > 0 >\$$ 
with probability 0.5 and $ 0>1 > \$$ with probability $0.5$.
Using multiple lexicographic orderings thus generated, we
order the suffixes obtained from $X_1$ and $X_2$, as shown in Figure~\ref{fig:sorting}.
We emphasize that the ``mask'' applied to obtain different orderings in Figure~\ref{fig:sorting} is applied to each suffix and not to the length-$n$ sequence $X_i$.
We note that with any ordering thus generated, 
the probability that the lexicographically first suffixes of $X_1$ and $X_2$ occur in their overlap is the same and approaches $\alpha$
as $n\to \infty$.
 
\begin{figure}[t] 
\centering
\ifarxiv
\includegraphics[width = 0.45\columnwidth]{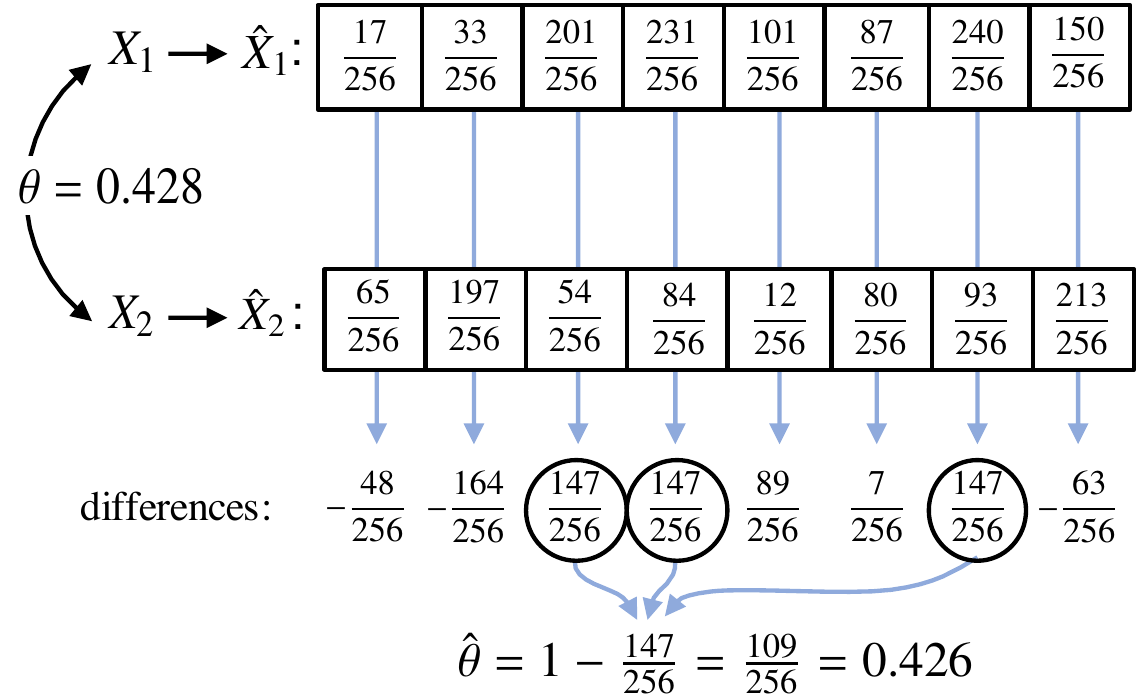}
\fi
\vspace{-0mm}
    \caption{
    Illustration of sketching and decoding scheme for $u = v = 8$.
    Each sketch $\hat X_i$ comprises $8$ numbers $[m_{i,j}]_8$, $j=1,...,u$, each with $8$-bit precision.
    Each of these numbers corresponds to the location of the lexicographically first suffix of $X_i$, each using a different lexicographic ordering.
    The decoder takes the difference between pairs of corresponding entries of $\hat X_1$ and $\hat X_2$, and then finds the mode of this set ($147/256$). 
    The estimate of $\theta$ is then $\hat \theta = 1 - 147/256$.
    The sketch size is $B =64$ bits.
    \label{fig:lochashing}}
\end{figure}

In order to build the improved sketch, we fix two parameters, $u$ and $v$.
We then generate $u$ different lexicographic orderings to compute $u$ different starting locations $m_{i,1},\dots,m_{i,u}$ for the lexicographically first suffix of $X_i$.
We build our sketch of $X_i$ by taking the $v$-bit truncation of $m_{i,j}/n$, for $j=1,\dots,u$; i.e.,
\al{
\Sk(X_i) = \left( 
[ m_{i,1}/n ]_v, \dots,
[ m_{i,u}/n ]_v 
\right).
}
This is illustrated in Figure~\ref{fig:lochashing}. 
Notice that this sketching procedure requires $B=uv$ bits.
For 
$j \in [1:u]$,
\aln{
1-([ m_{1,j}/n ]_v - [ m_{2,j}/n ]_v )^+
}
is an estimator for $\theta$ with the same performance as (\ref{eq:simpleloc}).
Moreover, in expectation, $\alpha u$ of them will satisfy
\al{
& m_{1,j} - m_{2,j} = (1-\theta) n 
\label{eq:shift2}
}
and thus 
\al{
[ m_{1,j}/n ]_v - [ m_{2,j}/n ]_v = (1-\theta) + \delta_j,
\label{eq:errorperentry}
}
where $\delta_j$ is an error term due to the finite-bit representation, and satisfies $|\delta_j| \leq 2\cdot 2^{-v}$.

Notice that, intuitively, for the remaining $(1-\alpha)u$ indices $j$, the value of $(m_{1,j}-m_{2,j})/n$ will be distributed in $(0,1)$ and is not expected to concentrate on any particular value.
Hence, an intuitive estimator for $\theta$ is obtained by 
letting $a$ be
the most frequent value (i.e., the mode) in the multiset
\aln{
\A = \{ [ m_{1,j}/n ]_v - [ m_{2,j}/n ]_v : 1 \leq j \leq u \},
}
and then setting $\hat \theta = 1-a$ if $a\geq 0$ and $\hat \theta = 0$ if $a < 0$.
This procedure is illustrated in Figure~\ref{fig:lochashing}.
We make one final modification to this estimator.
If the mode of $\A$ appears less than $\alpha_0 u /6$ times,
where $\alpha_0 \triangleq \theta_0/(2-\theta_0)$,
then we simply output $\hat \theta = 0$.
In Section~\ref{sec:proof}, 
we show that the resulting distortion satisfies
\al{
E & [(\theta-\hat \theta)^2 ] 
  \leq 2^{-2(v-1)} + (u+1) e^{-2u(\frac{\alpha_0}{12} - 2^{-(v-1)} )^2}.
 \label{eq:multipleloc}
}
It is easy to see that the expected distortion achieved with sketches of length $B = uv$ is significantly better than what is achieved with a single locational hash (\ref{eq:simpleloc}) and with the min-hash-based approach (\ref{eq:mhapbits}).
Setting $v = \alpha_0 \sqrt{B} /24$ and $u = B/v$ leads to the upper bound in Theorem~\ref{thm:main}.
\ifarxiv
Section~\ref{sec:proof} presents the ingredients to prove (\ref{eq:multipleloc}) and Theorem~\ref{thm:main}.
In the next section, we discuss how the performance of this approach is affected if we assume that the sequences $X_1$ and $X_2$ are independently corrupted by noise.
\fi

%% file: noisy.tex

\begin{figure}[b] 
\centering
\ifarxiv
\includegraphics[width = .5\columnwidth]{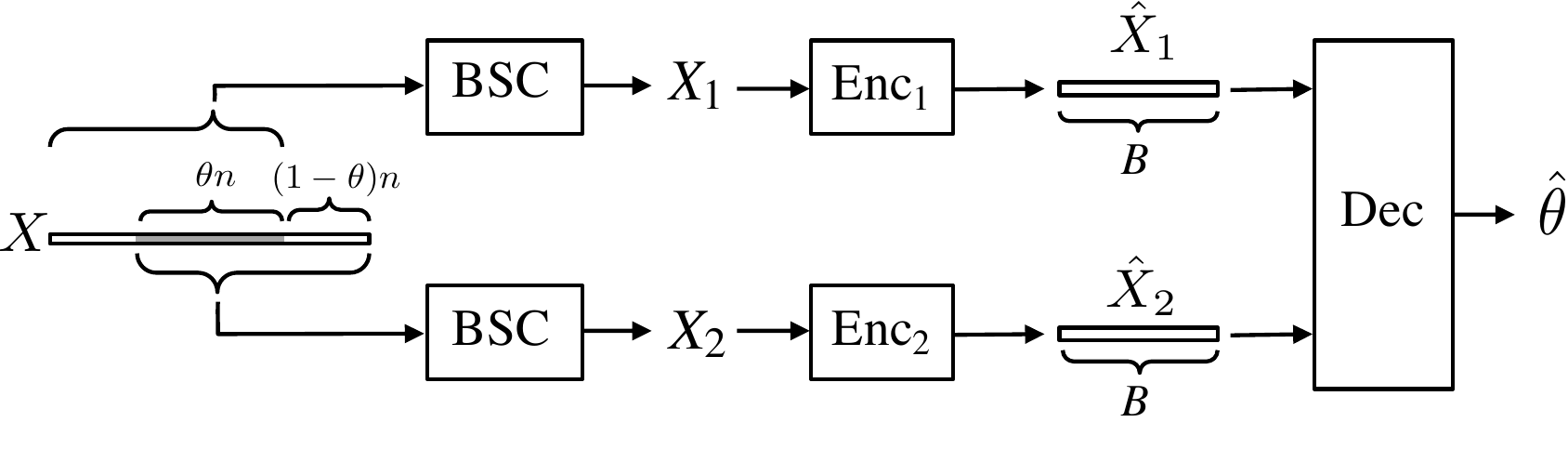}
\fi
    \caption{The source sequences $X_1$ and $X_2$ are individually encoded into $B$-bit sketches, from which the overlap $\theta$ must be estimated.
    \label{fig:sourcecoding-noisy}}
\vspace{-3mm}
\end{figure}

The absence of noise in 
our problem setup makes the approach and analysis presented in Section~\ref{sec:result} not directly practical.
In the context of genomics, for example, all DNA sequencing technologies have some error rate.
Hence, it is important to consider extensions of the 
formulation in Section~\ref{sec:problem}
to the case where the two sequences $X_1$ and $X_2$ are independently corrupted by noise.
A natural extension, shown in Figure~\ref{fig:sourcecoding-noisy}, can be obtained by assuming that $X_1$ and $X_2$ are generated by passing the length-$n$ prefix and suffix of $X$ through a binary symmetric channel (BSC) with crossover probability $p_n$, where $p_n \in [0,1]$ is allowed to vary with the blocklength $n$.

Now consider applying the locational hashing scheme from Section~\ref{sec:multipleloc} to this setting.
Consider the lexicographically first suffix of $X$, and suppose it occurs 
in the length-$\theta n$ overlap segment (gray segment in Figure~\ref{fig:sourcecoding-noisy}).
Notice that, if this suffix is corrupted by noise in $X_1$ (particularly if the noise occurs near the beginning of the suffix), then it is likely to no longer be the lexicographically first suffix of $X_1$.
This will affect the analysis of the distortion achieved in (\ref{eq:multipleloc}).


First we claim that, with high probability, the lexicographically first suffix of $X$ is determined by just considering the first $3 \log n$ bits of every suffix of $X$.
To see this, let $\Rs$ be the event that the underlying sequence $X$ has a repeat of length $k = 3 \log n$. 
By the union bound, $\Pr(\Rs)$ is at most
\al{
\sum_{i,j \in [1:n-k+1]} & \Pr\left( X[i:i+k-1] = X[j:j+k-1]  \right) \nonumber \\
& \leq n^2 \, 2^{-k} = 1/n. 
\label{eq:repeat0}
}
Hence, with probability $1-1/n$, the lexicographically first suffix of $X$ will be determined by the first $3 \log n$ bits of all suffixes (or fewer, if the suffix is close to the end of $X$).
Now consider the set $\S$ of length-$3 \log n$  substrings of $X_1$ and $X_2$ (after corruption by noise).
Notice that each substring that originates in the overlap segment of $X$ has a probability $\left((1-p_n)^{3 \log n}\right)^2$ of not being corrupted by noise in neither $X_1$ nor $X_2$.
Under the assumption that $\lim_{n \to \infty} p_n \log n = \beta$, for some $\beta > 0$,
\aln{
\lim_{n \to \infty} (1-p_n)^{6 \log n} 
= \lim_{n \to \infty} \left(1-\frac{\beta}{\log n}\right)^{6 \log n} = e^{-6 \beta}.
}
Hence, we expect that an $e^{-6 \beta}$ fraction of the length-$3 \log n$ substrings in the overlap are not corrupted by noise in $X_1$ and $X_2$ and 
\al{
\lim_{n \to \infty} \frac{|\S|}{n} \leq (2 - \theta e^{-6\beta} ).
\label{eq:sizes}
}
Since any of the strings in $\S$ is equally likely to be the lexicographically first, we conclude that the probability that an uncorrupted length-$3 \log n$ substring from the overlap is the lexicographically first in $\S$ is upper bounded by
\al{
\tilde \alpha \triangleq \frac{\theta e^{-6\beta}}{2 -\theta e^{-6\beta}}
\label{eq:tildealpha}
}
as $n \to \infty$.
Notice that this quantity corresponds to the parameter $\alpha$ defined in (\ref{eq:alpha}) for the error-free case (which corresponds to $\beta = 0$).
The parameter $\tilde \alpha$ is a lower bound to the probability that, for the $j$th  lexicographic ordering, 
(\ref{eq:shift2}) and (\ref{eq:errorperentry}) hold.
It is straightforward to see that the analysis in Section~\ref{sec:multipleloc} (and the technical details in Section~\ref{sec:proof}) hold with $\alpha$ replaced by $\tilde \alpha$ in the noisy case here considered.
Furthermore, since $\tilde \alpha \geq \left(2/(2-\theta)\right)e^{-6 \beta} = \alpha e^{-6 \beta} \geq \alpha_0 e^{-6 \beta}$,
we conclude that in the noisy case the distortion upper bound in (\ref{eq:multipleloc}) holds with $\alpha_0$ replaced by $\alpha_0 e^{-6 \beta}$, and we have the following generalization 
to Theorem~\ref{thm:main}.

\begin{theorem} \label{thm:noisy}
If the sequences $X_1$ and $X_2$ are independently corrupted by a BSC($p_n$) channel where $\lim_{n\to\infty} p_n \log n = \beta$,
the minimum sketch size $B$ required to achieve distortion $D$  satisfies
\al{
\frac18 \log \left(\frac{1-\theta_0}{D}\right) \leq B(D) 
\leq \frac{24^2 e^{12 \beta}}{\alpha_0^2} \ln^2 \left(\frac3D\right)
\label{eq:thmnoisy}
}
for any $D$ small enough.
\end{theorem}

Theorem~\ref{thm:noisy} provides an initial step to studying the sketching problem in the setting where the sequences are corrupted by noise.
In particular, it implies that if the error rate decays as $1/\log (n)$ (or faster), then it is still possible to achieve distortion $D$ with sketches of length $B = \log^2(1/D) \cdot O(1)$ as $n \to \infty$.
Furthermore, in the case where the error rate does not decay as $1/\log (n)$,
(e.g., when $p_n$ is constant or $p_n$ decays as $1/\log \log (n)$)
if we think of $\beta$ as $p_n \log n$, the upper bound in (\ref{eq:thmnoisy}) suggests that it is possible to achieve distortion $D$ with sketches of length
$\log^2(1/D) \cdot O(e^{12 \beta}) = \log^2(1/D) \cdot O(n^{(12 \ln 2 ) p_n})$.
Notice that this could still represent a significant reduction with respect to working with the entire sequences of length $n$.


%% file: proof.tex



In this section we provide the key steps for the achievability of Theorem~\ref{thm:main}.
\ifarxiv
Proofs of lemmas and the proof of the converse result are relegated to the Appendix.
\fi

The key step in the analysis of our sketching procedure and its decoder is obtaining, for the $j$th lexicographic ordering, $j \in [1:u]$, bounds for the pmf
\aln{
p_j(k) \triangleq \Pr \left( [m_{1,j}/n]_v - [m_{2,j}/n]_v = k \, 2^{-v} \right). 
}
Intuitively, this pmf has a probability mass of $\alpha$  near $k \approx (1-\theta)2^v$, and no other probability mass anywhere.
This is made precise in the following lemma.
\begin{lemma} \label{lem:pmf}
Let
$\K = \{ k : \  |k \, 2^{-v} - (1-\theta)| \leq 2^{-(v-1)}\}$ if $\theta > 0$ and $\K = \emptyset$ if $\theta = 0$. 
Then
\al{
\sum_{k \in \K} p_j(k) \geq \alpha - \ep_n,
\label{eq:lemeq1}
}
where $\lim_{n\to \infty} \ep_n = 0$.
Moreover,
\al{
\max \{ p_j(k) : k \notin \K \} \leq 
2^{-(v-1)}. \label{eq:lemeq2}
}
\end{lemma}
Given the lemma, we can bound the expected distortion as follows.
First, we note that the set $\K$ from Lemma~\ref{lem:pmf} has cardinality at most $5$ (as we work with a resolution of $2^{-v}$ and are considering
an interval of length $4 \times 2^{-v}$).
Hence, for $n$ large enough, there is at least one value of $k \in \K$ with $p_j(k) \geq \alpha/6$.
%
Notice that, 
since the $u$ different lexicographic orderings are chosen independently at random,
$(m_{1,j},m_{2,j})$ is independent of $(m_{1,\ell},m_{2,\ell})$ for $j \ne \ell$.
Hence, the random variables $[ m_{1,j}/n ]_v - [ m_{2,j}/n ]_v$, for $1 \leq j \leq u$, are independent (and identically distributed).

We bound the distortion by considering two separate cases.
If $\alpha = 0$ (which is equivalent to $\theta = 0$), then the estimator will be $\hat \theta = 0$ as long as no element of $\A$ appears more than $\alpha_0 u / 6$ times, and the distortion will be zero.
If this condition is violated for some element $k2^{-v}$ with $k \in \K$, then we incur a distortion of at most $(2^{-(v-1)})^2$.
If this condition is violated for some $k2^{-v}$ with $k \notin \K$, then the distortion can be at most $1$, but this happens with low probability, as shown in the following lemma (which holds regardless of whether $\alpha = 0$).
\begin{lemma} \label{lem:bins}
The probability that there is a value $k2^{-v}$ for $k \notin \K$ that appears more than $\alpha_0 u/12$ times in $\A$ is at most
\al{
\exp \left[ -2 u ( \alpha_0/12 - u 2^{-(v-1)})^2 \right],
\label{eq:alpha0u}
}
provided that $\alpha_0/12 > u 2^{-(v-1)}$.
\end{lemma}
Hence, when $\alpha = 0$ the expected distortion satisfies
\al{
E & [(\theta-\hat \theta)^2 ]  
 \leq 2^{-2(v-1)} + e^{-2u(\frac{\alpha_0}{12} - u 2^{-(v-1)} )^2}.
\label{eq:multipleloc1}
}
Next we consider the case $\alpha \geq \alpha_0$.
Notice that in this case Lemma~\ref{lem:bins} still holds.
Moreover, the probability that, for no $k \in \K$, $k2^{-v}$ appears at least $(\alpha_0/12)u$ times in $\A$ is at most
\al{
\exp \left( -2 u (\alpha/6-\alpha_0/12)^2 \right) \leq \exp \left( -2 u (\alpha_0/12)^2 \right).
}
We conclude that, if
$\alpha_0/12 > u2^{-(v-1)}$, 
the expected distortion in the case $\alpha \geq \alpha_0$ can bounded as 
\al{
E [(\theta-\hat \theta)^2 ]  
 & \leq 2^{-2(v-1)} + e^{-2u(\frac{\alpha_0}{12} - u2^{-(v-1)} )^2} + e^{-2u(\frac{\alpha_0}{12})^2} \nonumber \\
 & 
  \leq 2^{-2(v-1)} + 2 e^{-2u(\frac{\alpha_0}{12} - u 2^{-(v-1)} )^2}.
\label{eq:multipleloc2}
}
Since the right-hand side of (\ref{eq:multipleloc2}) is greater than the right-hand side of (\ref{eq:multipleloc1}), we can take (\ref{eq:multipleloc2}) to be a bound for both cases.

Finally, by setting $v = \alpha_0 \sqrt{B} /24$ $u = B/v$, we obtain 
\al{
E & [(\theta-\hat \theta)^2 ] 
\leq 3 \exp\left(-\sqrt{B} \alpha_0 / 24 \right),
}
which holds for $B$ large enough.
This implies that 
$
\ln\left( 3/D \right) \geq \alpha_0 \sqrt{B}/24,
$
which yields the bound in Theorem~\ref{thm:main}.
Since this bound only holds for $B$ large enough, 
our bound on $B(D)$ only holds for small enough $D$.

%% file: discussion.tex

We introduced an information-theoretic framework to study the problem of sketching sequences for pairwise alignment.
As a starting point, we considered the case where we have two binary sequences with an error-free overlap.
We showed that, for a desired distortion bound $D$, 
one can compress the strings into $B$-bit sketches 
with $B = \log^2(1/D)O(1)$.
Since the encoder produces a sketch with a time complexity that is linear in $n$ and the decoder produces an estimator $\hat \theta$ in time that is linear in the sketch size, the pairwise overlapping of a large number $m$ of length-$n$ sequences can be performed in time $O(mn) + O(m^2) \log^2(1/D)$, while existing methods would require $O(mn) + O(m^2) (1/D)$ for 
the same distortion.

Finally, we point out that, depending on the application, one may be interested in identifying local alignments beyond just overlaps (e.g., Figure~\ref{fig:alignment}(b)).
In these situations, the scheme we presented 
produces an estimate of the ``shift'' between the two sequences that leads to the largest match, rather than an estimate of the size of the match.
Modifying our scheme for the case of general overlaps is another  direction that we will consider in follow-up work.

\section*{Acknowledgements}
The research of I.~Shomorony was supported in part by NSF grant CCF-2007597 and NSF CAREER grant CCF-2046991.



%% file: appendix.tex
\subsection{Proof of Lemma~\ref{lem:pmf}} \label{sec:prooflemma}

To prove (\ref{eq:lemeq1}), we follow similar steps to those in Section~\ref{sec:simpleloc}, but more formally.
Let $\Rs$ be the event that the underlying sequence $X$ has a repeat of length $k = 3 \log n$, and $\bar \Rs$ its complement.
Then $\Pr(\Rs)$ is at most
\al{
\sum_{i,j \in [1:n-k+1]} & \Pr\left( X[i:i+k-1] = X[j:j+k-1]  \right) \\
& \leq n^2 \, 2^{-k} = 1/n. 
\label{eq:repeat}
}
Hence, with probability $1-1/n$, the lexicographically first suffix of $X$ will be determined by the first $3 \log n$ bits of all suffixes.
Hence, if the lexicographically first suffix of $X$ starts inside $X[(1-\theta)n+1:n-3\log n]$, it will be completely contained in the overlap between $X_1$ and $X_2$, and (\ref{eq:shift2}) will hold.
Therefore, as $n \to \infty$, 
Therefore, 
with probability approaching $\alpha$, (\ref{eq:errorperentry}) holds,
which in turn implies that 
\al{
[ m_{1,j}/n ]_v - [ m_{2,j}/n ]_v = k2^{-v}
\label{eq:proof1}
}
for some $k \in \K$.
This proves (\ref{eq:lemeq1}).

To prove (\ref{eq:lemeq2}), we define three segments of $X$, $S_1 = [1:(1-\theta)n - 3\log n]$,
$S_{12} = [(1-\theta)n : n - 3\log n]$, and 
$S_{2} = [n : (2-\theta)n  - 3\log n]$.
Notice that, as $n\to \infty$, we must have $m_{1,j} \in S_{1} \cup S_{12}$ and $m_{2,j} \in S_{2} \cup S_{12}$ with probability tending to $1$.
If $m_{1,j}$ and $m_{2,j}$ are both in $S_{12}$, they must correspond to the same suffix of $X$, and (\ref{eq:proof1})  will hold for $k \in \K$.
Conditioned on $\E \triangleq \bar \Rs \cap \{m_{1,j} \in S_{1}\} \cap \{m_{2,j} \in S_2\}$, $m_{1,j}$ and $m_{2,j}$ are  uniformly distributed in their intervals.
Hence, 
\aln{
\Pr & \left( [ m_{1,j}/n ]_v - [ m_{2,j}/n ]_v = k2^{-v} \middle| \E \right) \\ 
& \leq 
\max_\ell \Pr \left( [ m_{1,j}/n ]_v = \ell 2^{-v} \middle| \E \right) \leq \frac{2^{-v}}{1-\theta}.
}
Similarly, conditioning on the events 
$\E' \triangleq \bar \Rs \cap \{m_{1,j} \in S_{1}\} \cap \{m_{2,j} \in S_{12}\}$ and 
$\E'' \triangleq \bar \Rs \cap \{m_{1,j} \in S_{12}\} \cap \{m_{2,j} \in S_{2}\}$, we see that the conditional pmf is always upper bounded by $2^{-v}/(1-\theta)$.
Since $\Pr(\E \cup \E' \cup \E'') \to (1-\alpha)$ as 
$n \to \infty$,  
we conclude that
\aln{
\max \{ p_j(k) : k \notin \K \} & \leq \left(\frac{1-\alpha}{1-\theta}\right) 2^{-v}
= \left(\frac{2}{2-\theta}\right) 2^{-v},
}
and since $\theta \in (0,1)$, (\ref{eq:lemeq2}) is proved.

\subsection{Lower Bound on Sketch Size} \label{sec:converse}

In this section we provide the converse result to establish the lower bound in Theorem~\ref{thm:main}; i.e., $B \geq \tfrac18 \log \left( \frac{1-\theta_0}{D}\right)$.
First we observe that, by combining both sketches we have a total of $2B$ bits, which implies that the output of any decoder function $\Dec(\cdot)$ can take at most $2^{2B}$ values in 
$(0,1)$.
Hence, for any decoder, there exists $\theta \in \{0\} \cup [\theta_0,1]$ that is at a distance $(1-\theta_0) 2^{-4B}$ from every value in the range of $\Dec(\cdot)$.
Hence,
\aln{
\max_{\theta} E[ (\theta - \hat \theta)^2 ] 
\geq (1-\theta_0) 2^{-8B}.
}
This, in turn, implies that $D \geq (1-\theta_0) 2^{-8B}$.

\subsection{Proof of Lemma \ref{lem:bins}}

First we consider two strings $X_1$ and
$X_2$ with $\alpha = 0$.
In order to simplify the notation, we define the random variables 
\aln{
Y_i = (\Sk(X_1)-\Sk(X_2))_i \cdot 2^v,
}
for $i=1,\dots,u$.
The scaling by $2^v$ implies that $Y_i \in \{-2^v, \dots, 0, \dots, 2^v-1\}$ for $i=1,\dots,u$.
From the discussion in Section~\ref{sec:proof}, $Y_1,\dots,Y_u$ are i.i.d.~random variables.
We also define $\mathcal{U}_i$ to be the set of the $u$ most 
seen values in $\{ 0, \dots, 2^v-1\}$ 
among $Y_1,\dots,Y_i$.
In case of ties, we pick 
the numerically smaller value to be in $\mathcal{U}_i$. 
Notice that, following this rule, 
$\mathcal{U}_0 = \{0,1,\dots,u\}$.
We also define random variables $\Gamma_i$, $i=1,...,u$, as
\begin{align}
    \Gamma_i &= \begin{cases}
    1 & Y_i \in \mathcal{U}_{i-1}\\
    0 & \text{otherwise}
    \end{cases}
\end{align}
and we let $\Gamma_0 = 0$.
Intuitively, $\Gamma_i$ is an indicator for the ``collision'' event that $Y_i$ attains the same value of one of the previously seen values in $\{Y_1,\dots,Y_{i-1}\}$.
Since the random variables $Y_1,\dots,Y_{u}$ are i.i.d., given $Y_1,\dots,Y_{i-1}$, the distribution of $Y_i$ does not change, but the distribution of $\Gamma_i$ does.
More precisely, since $Y_1,\dots,Y_{i-1}$ fully determines $\U_{i-1}$, we have that
\al{
E \left[ \Gamma_i  \middle| Y_1,\dots,Y_{i-1} \right] 
& = \sum_{k \in \U_{i-1}} \Pr( Y_i = k) \leq u 2^{-{(v-1)}}.
\label{eq:martingale1}
}
The inequality follows from Lemma~\ref{lem:pmf}, since, when $\alpha = 0$, $\theta = 0$ and $\K = \emptyset$.
If we define $\tilde{\Gamma}_i = {\Gamma}_i - u2^{-(v-1)}$,
we have
\aln{
E \left[ \tilde{\Gamma}_i  \middle| Y_1,\dots,Y_{i-1} \right] \leq 0
}
for $i=1,\dots,u$, which implies that $\tilde{\Gamma}_1,\dots,\tilde{\Gamma}_u$ is a supermartingale difference sequence (i.e., $\sum_{j=1}^i \tilde{\Gamma}_j$, $i=1,\dots,u$ is a supermartingale with respect to $Y_1,...,Y_u$.
Moreover, since $0 < \tilde{\Gamma}_i < 1$, by
Azuma-Hoeffding's inequality, for any $\lambda > 0$,
\aln{
\Pr\left(\sum_{i=1}^u \tilde{\Gamma}_i  \ge \lambda u \right) 
\leq \exp \left( -\frac{2 u^2 \lambda^2}{\sum_{i=1}^u 1^2 } \right)
= \exp \left( - 2 u \lambda^2 \right).
}
Now let $F$ be the maximum number of times 
we see an element in the vector $\Sk(X_1)-\Sk(X_2)$ (or in the multi-set $\{Y_1,\dots,Y_u\}$). 
We then have that $\sum_{i=1}^u \Gamma_i \ge F-1$.
Further we have that, if $Y_i =j$ for some $j$,
then $j \in \mathcal{U}_k, k\ge i$, because we at most see $u$ distinct items in $Y_1,\dots,Y_u$.
We conclude that 
\begin{align}
    \Pr\left[F \ge \frac{\alpha_0 u}{12}\right] &\le  \Pr\left[\sum_{i=1}^u \Gamma_i \ge \frac{\alpha_0 u}{12} + 1 \right] \\
    &\le \Pr\left[\sum_{i=1}^u \Gamma_i \ge \frac{\alpha_0 u}{12}  \right] \\
    &= \Pr\left[\sum_{i=1}^u (\Gamma_i - u 2^{-(v-1)}) \ge \frac{\alpha_0 u}{12} - u^2 2^{-(v-1)}  \right] \\
    &= \Pr\left[\sum_{i=1}^u \tilde{\Gamma}_i  \ge u\left( \frac{\alpha_0}{12} - u 2^{-(v-1)} \right) \right] \\
    &\overset{(i)}{\le} \exp \left( -2u \left(\frac{\alpha_0 }{12} - u 2^{-(v-1)}\right)^2 \right),
\end{align}
where $(i)$ follows from Azuma-Hoeffding's inequality.

Finally, we point out that, in the case where $\alpha > 0$ (which implies $\alpha \geq \alpha_0$), the same argument holds under minor modifications.
In particular, we only need to consider values of $k \in \{0,\dots,2^v\} - \K$ in the definition of $\U_i$, since we are only interested in the event that one of those values appears many times.
The rest of the proof follows in a straightforward manner.